\documentclass[3p,times]{elsarticle}

\usepackage{ecrc}
\usepackage{hyperref}

\volume{00}

\firstpage{1}

\journalname{Nuclear Physics A}

\runauth{David d'Enterria and Alexander Snigirev}

\jid{npa}

\jnltitlelogo{Nuclear Physics A}

\usepackage{amssymb}

\biboptions{square,comma,numbers,sort&compress}

\usepackage[figuresright]{rotating}

\usepackage{multirow}

\newcommand{\sqrtsnn}{\sqrt{s_{_{\mbox{\rm \tiny{NN}}}}}}
\newcommand{\alphaS}{\alpha_{\rm s}}

\newcommand{\LumiInt}{\mathcal{L}_{\mbox{\rm \tiny{int}}}}

\newcommand{\pp}{{\rm{p-p}}}

\newcommand{\pA}{{\rm{p-A}}}
\newcommand{\pN}{{\rm{p-N}}}
\newcommand{\pPb}{{\rm{p-Pb}}}
\newcommand{\NN}{{\rm{N-N}}}
\newcommand{\AaAa}{{\rm{A-A}}}
\newcommand{\PbPb}{{\rm{Pb-Pb}}}

\newcommand{\TpA}{\rm T_{_{\rm pA}}}

\newcommand{\TpAsq}{\rm T^{2}_{_{\rm pA}}}

\newcommand{\dtwor}{{\rm{d^2r}}}

\newcommand{\mcfm}{{\sc mcfm}}

\newcommand{\sigmaDPS}{\sigma^{{\rm {\tiny DPS}}}}

\newcommand{\sigmaSPS}{\sigma^{{\rm {\tiny SPS}}}}
\newcommand{\sigmaeff}{\sigma_{\rm eff}}
\newcommand{\sigmaeffpp}{\sigma_{_{\rm eff,pp}}}
\newcommand{\sigmaeffpA}{\sigma_{_{\rm eff,pA}}}
\newcommand{\sigmaeffAA}{\sigma_{_{\rm eff,AA}}}
\newcommand{\sigmaDPSjpsijpsi}{\sigma^{{\rm {\tiny DPS}}}_{{\jpsi\jpsi}}}

\newcommand{\NDPS}{\rm N^{{\rm {\tiny DPS}}}}

\newcommand{\jpsi}{J/\psi}

\def\order#1{\mathcal{O}{(#1)}}

\begin{document}

\begin{frontmatter}



\dochead{}

\title{Pair production of quarkonia and electroweak bosons from\\ 
double-parton scatterings in nuclear collisions at the LHC}


\author[cern]{\underline{David d'Enterria}} 
\address[cern]{CERN, PH Department, 1211 Geneva, Switzerland}
\author[msu]{Alexander~M.~Snigirev}
\address[msu]{Skobeltsyn Institute of Nuclear Physics, Moscow State University, 119991 Moscow, Russia}

\begin{abstract}
Cross sections for the concurrent production of pairs of quarkonia ($\jpsi$,$\Upsilon$) and/or W,Z
gauge bosons from double-parton scatterings (DPS) in high-energy proton-nucleus and
nucleus-nucleus collisions at the LHC are calculated. The estimates are based on next-to-leading-order
perturbative QCD predictions, including nuclear modifications of the parton densities, 
for the corresponding single-scattering cross sections. 
Expected event rates for $\jpsi+\jpsi$, $\jpsi+\Upsilon$, $\jpsi+$W, $\jpsi+$Z, $\Upsilon+\Upsilon$, 
$\Upsilon+$W, $\Upsilon+$Z, and same-sign W+W production in their (di)leptonic decay modes,
after typical acceptance and efficiency losses, are given for \pPb\ and \PbPb\ collisions. 
\end{abstract}


\end{frontmatter}


\section{Introduction}
\label{sec:intro}

Particle production in high-energy hadron-hadron collisions is dominated by multiple interactions of their
constituent partons. At LHC energies, most of the quarks and gluons forming the colliding protons and nuclei
interact at semi-hard scales $\order{1-3\rm~GeV}$ and fragment into ``minijet'' bunches of
final-state hadrons. About half of the particles produced in a standard proton-proton (\pp), proton-nucleus
(\pA), and nucleus-nucleus (\AaAa) collision at the LHC come from the radiation and fragmentation of such
secondary partonic interactions (the other half coming from the hardest parton-parton scattering in the event, 
and/or from softer ``peripheral'' interactions). Many basic event properties --such as the
distributions of hadron multiplicities in ``minimum bias'' collisions as well as the underlying event activity
in hard scattering interactions-- can only be explained by taking into account hadron production issuing 
from such multiparton interactions (MPIs) occurring in each single \pp, \pA, and \AaAa\
collision~\cite{Bartalini:2010su,Bartalini:2011jp}. Monte Carlo (MC) generators reproduce all such properties
by modeling the colliding hadrons through longitudinal parton distribution functions (PDFs) complemented with a
parametrisation of their transverse parton profile as a function of impact-parameter (${\bf b}$).

The existence of MPIs at semi-hard scales naturally supports the possibility of double-parton scatterings
(DPS) producing, in the same collision, two independently-identified particles at {\it harder} scales, $\order{3-100\rm~GeV}$. 
Various differential distributions in W+jets~\cite{Aad:2013bjm,Chatrchyan:2013xxa} and 
$\jpsi$+W~\cite{Aad:2014xca} processes in \pp\ at the LHC show excesses of events --above the expectations
from single-parton scatterings (SPS) alone-- consistent with DPS contributions. 
There are, however, still large uncertainties on the DPS extraction due to (i) the contributions of
higher-order SPS processes, (ii) our limited knowledge of the proton transverse parton
profile~\cite{Diehl:2011yj} and its energy evolution, and (iii) the role of multi-parton correlations in the
hadronic wave functions~\cite{Calucci:2010wg}. In Refs.~\cite{d'Enterria:2012qx,d'Enterria:2013ck}, we have
highlighted the importance of studying DPS in \pA\ and \AaAa\ collisions as a complementary means to clarify
such open issues.\\
 
In a generic (model-independent) way, one can write the DPS cross section simply as the product of the
SPS cross sections, $\sigmaSPS$, normalised by an effective cross section $\sigmaeffpp$ characterising the
transverse area of the hard partonic interactions i.e., in the case of \pp\ collisions,
\begin{equation} 
\sigmaDPS_{(pp\to a b)} = \left(\frac{m}{2}\right) \frac{\sigmaSPS_{(pp\to a)} \cdot \sigmaSPS_{(pp\to b)}}{\sigmaeffpp}\,,
\label{eq:sigmaDPSpp}
\end{equation}
where $\sigmaSPS$ is computable perturbatively, at a given order of accuracy in 
$\alphaS$,
through convolution of the partonic subprocess cross sections and the proton PDFs, and $(m/2)$ is a 
combinatorial factor accounting for (in)distinguishable $m=2$ ($m=1$) final-states.
A numerical value $\sigmaeffpp\approx$~15~mb has been obtained from empirical fits to 
DPS-sensitive distributions in \pp\ at the LHC~\cite{Aad:2013bjm,Chatrchyan:2013xxa}.
One can identify $\sigmaeffpp$ with the inverse of the proton overlap-function squared: 
$\sigmaeffpp = \left[ \int d^2b \, t^2({\bf b})\right]^{-1}$, under the two following 
assumptions: (i) the (generalised) proton PDFs can be decomposed into longitudinal  
and transverse components, with the latter expressed in terms of the overlap function 
$t({\bf b}) = \int f({\bf b_1}) f({\bf b_1 -b})d^2b_1 $ for a given parton transverse thickness 
function $f({\bf b})$, and
(ii) the longitudinal component reduces to the ``diagonal'' product of two independent
single PDFs. Explaining the fact that the measured $\sigmaeffpp$ is about a factor of two smaller than
estimates based on naive geometric parametrisations of the proton profile~\cite{Abe:1997xk}, and ascertaining
the evolution of $\sigmaeffpp$ with collision energy, remain two important open issues in DPS studies.\\ 

In \pA\ collisions, the SPS cross section is simply that of proton-nucleon (\pN) collisions --taking into account
possible modifications of the nuclear PDFs compared to the free nucleon-- scaled by the number of nucleons
(A) in the nucleus~\cite{d'Enterria:2003qs}. The DPS cross sections are further
enhanced due to interactions where the two partons of the nucleus belong either to the
same nucleon or to two different nucleons~\cite{Strikman:2001gz}. The corresponding ``pocket formula'' 
reads~\cite{d'Enterria:2012qx}:
\begin{eqnarray} 
\sigmaDPS_{pA\to a b} = \left(\frac{m}{2}\right) \frac{\sigmaSPS_{pN \to a} \cdot \sigmaSPS_{pN \to b}}{\sigmaeffpA}\,,
\mbox{ with } \; \sigmaeffpA = \frac{\sigmaeffpp}{A+\sigmaeffpp\,\rm{F}_{pA}} \approx 22.6 \mbox{ $\mu$b},
\label{eq:sigmapADPS}
\end{eqnarray} 
with $\rm{F}_{pA} = \frac{A-1}{A} \int \TpAsq({\bf r})\,\dtwor$, where $\TpA({\bf r})$ is the standard Glauber
nuclear thickness function~\cite{d'Enterria:2003qs}, and the last numerical equality holds for \pPb\ using 
A~=~208, $\sigmaeffpp =$~15~mb, and $\rm{F}_{pA}$~=~30.4~mb$^{-1}$. The DPS cross sections in \pPb\ are thus
enhanced by a factor of $\sigmaeffpp/\sigmaeffpA \approx 3\,A \approx$~600 compared to \pp. 
Since the parameter F$_{pA}$ depends on the (comparatively better) known transverse density profile of nuclei,
one can exploit such a large DPS signal in \pA\ collisions to determine the value of $\sigmaeffpp$
independently of \pp\ measurements.\\ 

In the \AaAa\ case, the single-parton cross section is that of \pp\ (again, modulo nuclear PDF modifications)
scaled by A$^2$, and the DPS cross section is further enhanced due to contributions coming from interactions
where the two partons belong or not to the same pair of nucleons of the colliding nuclei~\cite{d'Enterria:2013ck}:
\begin{eqnarray} 
\sigmaDPS_{(AA\to a b)} = \left(\frac{m}{2}\right) \frac{\sigmaSPS_{(NN \to a)} \cdot \sigmaSPS_{(NN \to b)}}{\sigmaeffAA},
\mbox{ with } \; \sigma_{\rm eff,AA} = \frac{1}{A^2\left[\sigmaeffpp^{-1}+\frac{2}{A}\,\rm{T}_{AA}(0)\,+\,\frac{1}{2}\,\rm{T}_{\rm AA}(0)\right]} \approx 1.5 \mbox{ nb} \,.
\label{eq:sigmaAADPS}
\end{eqnarray}
Here $\rm{T}_{AA}({\bf b})$ is the nuclear overlap function~\cite{d'Enterria:2003qs} and
the last equality holds for \PbPb\ ($\rm{T}_{AA}(0)$~=~30.4~mb$^{-1}$).
The three terms in the denominator indicate the three DPS contributions from interactions where:
(i) the two colliding partons belong to the same pair of nucleons, (ii) partons from one nucleon in one
nucleus collide with partons from two different nucleons in the other nucleus, and (iii) the two colliding
partons belong to two different nucleons from both nuclei. 
Their relative contributions are approximately 1:4:200, with the last (Glauber binary scaling) term clearly
dominating. Whereas the single-parton cross sections in \PbPb\ are enhanced by a factor of A$^2~\simeq~4\cdot
10^4$ compared to that in \pp, the corresponding double-parton cross sections are boosted by a much larger
factor of $\sigmaeffpp\,/\sigma_{\rm eff,AA}\propto A^{3.3}/5 \simeq 9 \cdot 10^6$.
Pair-production of pQCD probes issuing from DPS represents thus an important feature of heavy-ion collisions at the
LHC and needs to be taken into account in any attempt to interpret the event-by-event characteristics
of any observed suppression and/or enhancement of their yields in \PbPb\ compared to \pp\ data.

\section{Production of $\jpsi\jpsi$ in Pb-Pb at 5.5 TeV and W$^\pm$W$^\pm$ in p-Pb at 8.8 TeV via double parton
  scatterings}
\label{sec:ppb}

\begin{figure}[hbtp!]
  \centering
  \includegraphics[width=0.47\columnwidth,height=8.3cm]{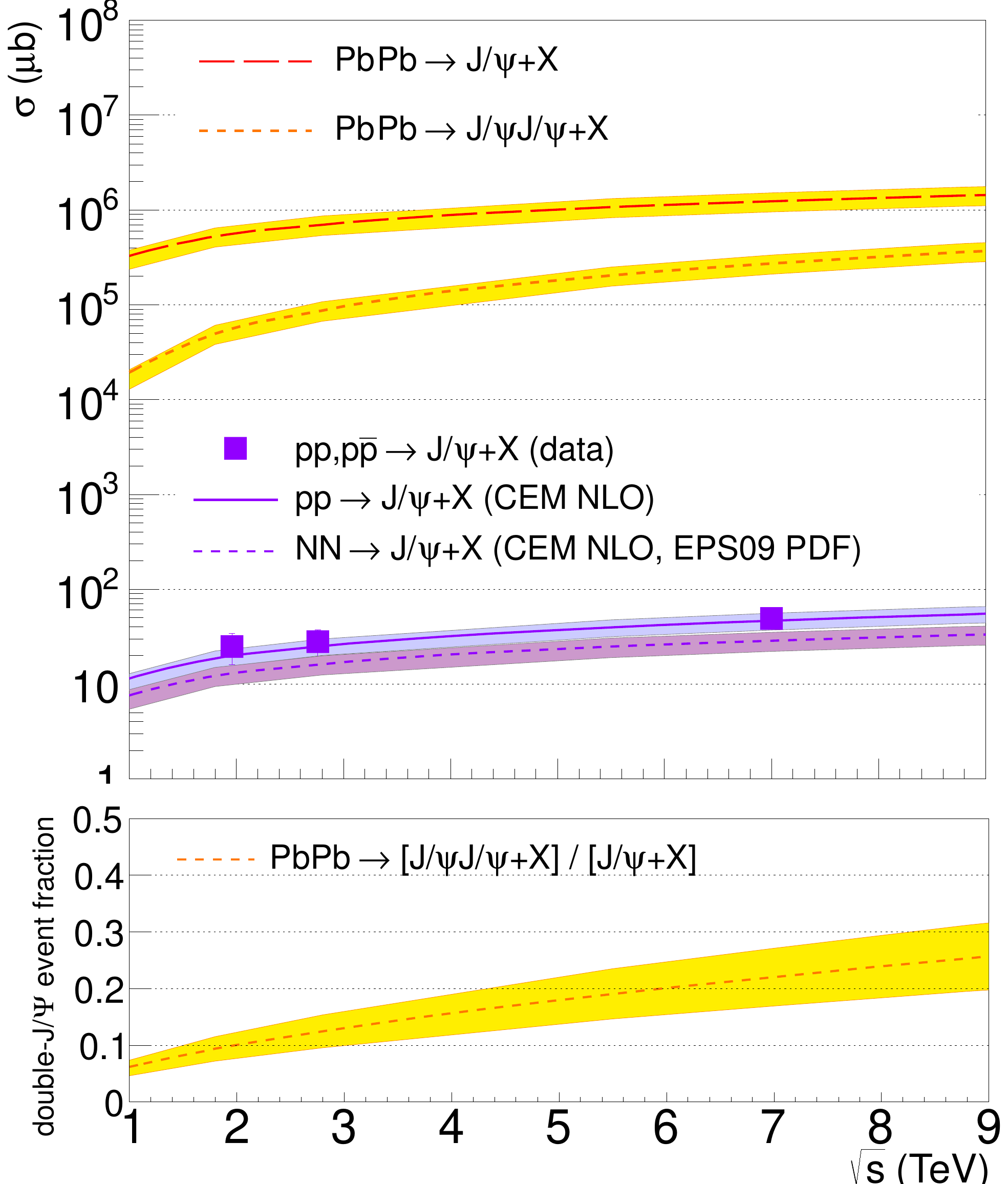}
  \hspace{0.2cm}
  \includegraphics[width=0.49\columnwidth,height=8.3cm]{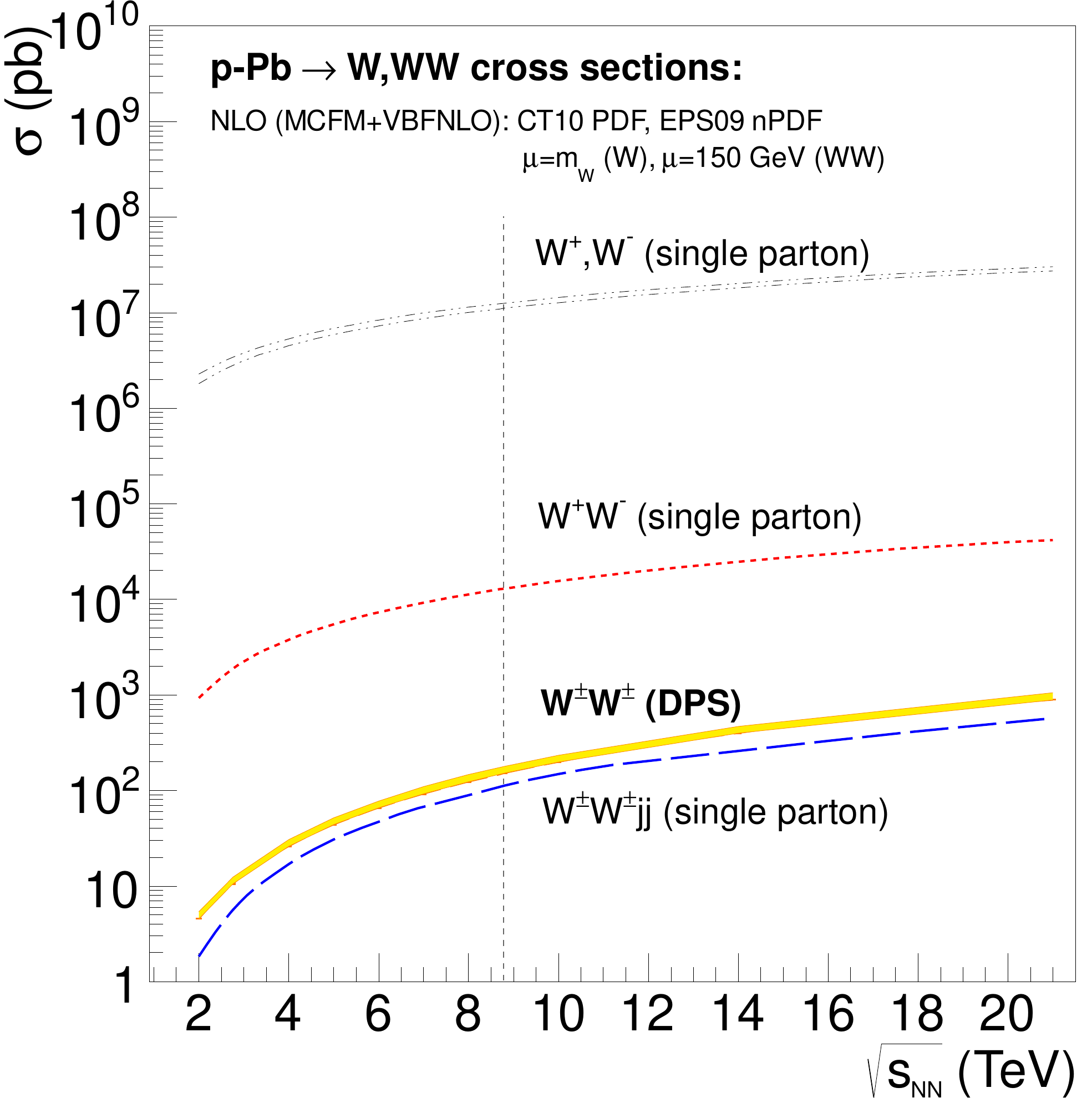} 
  \caption{Cross sections as a function of c.m. energy for:
  (i) prompt-$\jpsi$ production in \pp, \NN, and \PbPb\ collisions and for double-parton $\jpsi\jpsi$ in
  \PbPb~\cite{d'Enterria:2013ck} (left); and 
  (ii) single-W 
  and W-pair boson(s) from single-parton and double-parton scatterings in \pPb\ (right)~\cite{d'Enterria:2012qx}.} 
  \label{fig:sigmaDPS_vs_sqrts}
\end{figure}

The DPS cross section for double-$\jpsi$ production in \PbPb\ has been computed with
Eq.~(\ref{eq:sigmaAADPS}) using the colour evaporation model ({\sc cem}) NLO 
predictions~\cite{Vogt:2012vr} for the SPS $\jpsi$ cross section, 
which reproduce well the experimental data (squares in Fig.~\ref{fig:sigmaDPS_vs_sqrts} left),
including EPS09 nuclear PDFs~\cite{Eskola:2009uj}.
The two top curves in Fig.~\ref{fig:sigmaDPS_vs_sqrts} (left) show the single-$\jpsi$ (dashes) and
double-$\jpsi$ (dots) cross sections in \PbPb\ versus nucleon-nucleon c.m. energy 
$\sqrtsnn$. Their ratio is shown in the bottom panel.  
At the nominal \PbPb\ energy of 5.5~TeV, the single prompt-$\jpsi$ cross sections is $\sim$1~b, and $\sim$20\%
of such collisions are accompanied by the production of a second $\jpsi$ from a double parton interaction. 
The probability of $\jpsi$-$\jpsi$ DPS production increases rapidly with decreasing impact-parameter and
$\sim$35\% of the the most central \PbPb$\,\to\jpsi+X$ collisions have a second $\jpsi$ in the final
state~\cite{d'Enterria:2013ck}. Accounting for dilepton decays, acceptance and efficiency --which result in a
$\sim$3$\cdot$10$^{-7}$ reduction factor in the ATLAS/CMS (central) and ALICE (forward) rapidities-- the
visible cross section is $d\sigmaDPSjpsijpsi/dy|_{y=0,2} \approx$~60~nb, i.e. about 250 double-$\jpsi$ events
per unit-rapidity in the four combinations of dielectron and dimuon channels for a $\LumiInt$~=~1~nb$^{-1}$
integrated luminosity, assuming no net in-medium $\jpsi$ suppression/enhancement.
These results show quantitatively that the observation of a $\jpsi$ pair in a given \PbPb\
event should not be (blindly) interpreted as indicative of $\jpsi$ production via $\rm c \bar{c}$ regeneration in
the quark-gluon-plasma~\cite{Andronic:2010dt}, since DPS constitute an important component of the total $\jpsi$ yield
with or without final-state dense medium effects.


Same-sign WW production --whose theoretical cross section has small uncertainties and its experimental
signature is characterised by a ``clean'' final-state with like-sign leptons plus large missing transverse energy 
from the undetected neutrinos-- has no SPS backgrounds at the same order in $\alphaS$, and has been proposed
since long as a ``smoking gun'' of DPS in \pp\ collisions~\cite{Kulesza:1999zh}. We have computed the DPS cross
section in \pPb\ via Eq.~(\ref{eq:sigmapADPS}) 
using \mcfm~6.2~\cite{Campbell:2011bn} at NLO accuracy for the single-parton W cross section, 
with CT10 proton~\cite{Lai:2010vv} and EPS09 nuclear~\cite{Eskola:2009uj} PDFs and 
theoretical scales $\mu = \mu_F = \mu_R$~=~$m_W$.
Figure~\ref{fig:sigmaDPS_vs_sqrts} (right) shows the total cross sections for all relevant 
processes in the range $\sqrtsnn \approx$~~2--20~TeV. 
At the nominal 8.8~TeV, the same-sign WW DPS cross section is $\sigmaDPS_{pPb\to WW}\approx$~150~pb (yellow
thick curve), i.e. a factor of 1.5 times higher than the SPS background, $\sigmaSPS_{pPb\to WWjj}$, 
obtained adding the QCD and electroweak 
cross sections for the production of
W$^+$W$^+$ (W$^-$W$^-$) plus 2 jets pairs (lowest dashed curve). Accounting for the leptonic decay ratios and
applying standard ATLAS/CMS 
acceptance and reconstruction cuts, one expects about 10 DPS same-sign WW
events in 2~pb$^{-1}$ integrated luminosity~\cite{d'Enterria:2012qx}.

\section{Pair production of quarkonia and vector bosons in double-parton scatterings in p-Pb and Pb-Pb }
\label{sec:compilation}

Table~\ref{tab:1} collects the DPS cross sections for quarkonium and electroweak-boson pair production in
\pPb\ and \PbPb\ collisions at the LHC obtained via Eqs.~(\ref{eq:sigmapADPS}) and (\ref{eq:sigmaAADPS})
respectively. The corresponding NLO SPS cross sections (computed with {\sc cem} for $\jpsi,\Upsilon$; and
\mcfm\ for W,Z) are: (i) $\sigmaSPS_{NN\to\jpsi}$~=~25~$\mu$b, $\sigmaSPS_{NN\to\Upsilon}$~=~1.7~$\mu$b, 
$\sigmaSPS_{NN\to W}$~=~30~nb, $\sigmaSPS_{NN\to Z}$~=~20~nb at 5.5~TeV; and
(ii) $\sigmaSPS_{pN\to\jpsi}$~=~45~$\mu$b, $\sigmaSPS_{pN\to\Upsilon}$~=~2.6~$\mu$b, 
$\sigmaSPS_{pN\to W}$~=~60~nb, and $\sigmaSPS_{pN\to Z}$~=~35~nb at 8.8~TeV. 
The visible DPS yields for $\cal{L}_{\rm int}$~=~1~pb$^{-1}$ and 1~nb$^{-1}$, are obtained 
taking into account: BR($\jpsi,\Upsilon$,W,Z)~=~6\%, 2.5\%, 11\%, 3.4\% per (di)lepton decay; 
plus simplified acceptance and efficiency losses: ${\cal A\times E}(\jpsi$)~$\approx$~0.01 (over $|y|=0,2$),
and ${\cal A\times E}(\Upsilon;W,Z)\approx$~0.2, 0.5 (over $|y|<2.5$). All listed processes 
are in principle observable in the LHC heavy-ion runs. Other DPS processes like W+Z and Z+Z have much 
lower visible cross sections and are not quoted. 

\begin{table}[htbp]
\renewcommand{\arraystretch}{1.3}
\begin{center}
\begin{tabular}{l|lcccccccc}\hline
System & & $\jpsi+\jpsi$ & $\jpsi+\Upsilon$ & $\jpsi$+W & $\jpsi$+Z & $\Upsilon+\Upsilon$ & $\Upsilon+$W & $\Upsilon+$Z & ss\,WW \\\hline
\PbPb & $\sigmaDPS$            & 210 mb & 28 mb & 500 $\mu$b & 330 $\mu$b & 960 $\mu$b & 34 $\mu$b & 23 $\mu$b & 630 nb \\
5.5 TeV & $\NDPS$ (1 nb$^{-1}$)& $\sim$250 & $\sim$340 & $\sim$65 & $\sim$14 & $\sim$95 & $\sim$35 & $\sim$8 & $\sim$15 \\\hline
\pPb & $\sigmaDPS$             & 45 $\mu$b &  5.2 $\mu$b & 120 nb & 70 nb & 150 nb & 7 nb & 4 nb & 150 pb \\
8.8 TeV & $\NDPS$ (1 pb$^{-1}$)& $\sim$65 & $\sim$60 & $\sim$15 & $\sim$3 & $\sim$15 & $\sim$8 & $\sim$1.5 & $\sim$4 \\\hline
\end{tabular}
\caption{DPS production cross sections of double-$\jpsi$, $\jpsi+\Upsilon$, $\jpsi$+W, $\jpsi$+Z,
  double-$\Upsilon$, $\Upsilon$+W, $\Upsilon$+Z, and  same-sign WW in \PbPb\ and \pPb\ at the LHC. The
  corresponding DPS yields, after (di)lepton decays and acceptance+efficiency losses, are given for
  1~nb$^{-1}$ and 1~pb$^{-1}$ respectively.}  
\label{tab:1}
\vspace{-0.5cm}
\end{center}
\end{table}

\section{Summary}

Multiparton interactions are a major contributor to particle production in hadronic collisions at
high energy. The large transverse parton density in nuclei results in a high probability of having 
two truly hard scatterings in \pA\ and \AaAa\ collisions. The cross sections for the 
simultaneous production of quarkonia and/or electroweak bosons from DPS processes in nuclear collisions at the
LHC, have been computed using NLO predictions for the corresponding single-parton cross sections. Processes
such as double-$\jpsi$, 
$\jpsi\,\Upsilon$, $\jpsi$\,W, $\jpsi\,$Z, double-$\Upsilon$, $\Upsilon\,$W, $\Upsilon\,$Z, and same-sign W\,W
production have large cross sections and visible event rates for the nominal LHC luminosities. The study of
such processes in \pPb\ can help determine the effective $\sigmaeff$ parameter characterising the transverse
parton distribution in the nucleon. Double-$\jpsi$ and double-$\Upsilon$ production in \PbPb\ provide
interesting insights on the event-by-event dynamics of quarkonia in hot and dense strongly-interacting matter. 

\paragraph*{\bf Acknowledgments}
We are grateful to Ramona~Vogt for providing the {\sc cem} predictions. This work is partly supported by
a joint research grant CERN-Russian Foundation of Basic Research No.~12-02-91505.



\bibliographystyle{elsarticle-num}
\bibliography{dde_proceeds_qm14}







\end{document}